\documentclass[peerreview]{IEEEtran}

\usepackage{subcaption}

\usepackage[utf8]{inputenc}
\usepackage{amsmath,amssymb}

\usepackage{graphicx}
\usepackage{grffile}

\usepackage{color}
\usepackage{array}

\usepackage{algorithm}
\usepackage{algpseudocode}

\usepackage[colorinlistoftodos,prependcaption,textsize=small]{todonotes}
\usepackage{blindtext}
\presetkeys%
    {todonotes}%
    {inline,backgroundcolor=yellow}{}

\usepackage[section]{placeins}

\begin{document}

\title{Clustering of Musical Pieces through Complex Networks: an Assessment over Guitar Solos}
\author{\IEEEauthorblockN{Stefano Ferretti}\\
\IEEEauthorblockA{Department of Computer Science and Engineering, University of Bologna\\
Bologna, Italy\\
s.ferretti@unibo.it}
}

\maketitle
\begin{abstract}
Musical pieces can be modeled as complex networks. This fosters innovative ways to categorize music, paving the way towards novel applications in multimedia domains, such as music didactics, multimedia entertainment and digital music generation. Clustering these networks through their main metrics allows grouping similar musical tracks. To show the viability of the approach, we provide results on a dataset of guitar solos.
\end{abstract}



\section{Introduction}

As per many other real world problems, data, phenomena and systems, also music can be modeled through complex network theory \cite{icme16,Ferretti2017271,Liu2010126}.
Music received a lot of attention in many computing domains, with particular focus on 
indexing, classification, clustering, summarization. However, the use of complex systems and network theory to characterize music has been introduced only recently.
Categorization, recommendation and classification are main use cases that have been explored in this sense \cite{Angeler2016,MMUL.2017.4,Oord:2013,Patra:2013,thickstun2016learning}.
Moreover, some other seminal works have been recently proposed where music, expressed in terms of score sheets, is treated as a complex network
\cite{icme16,Ferretti2017271,Liu2010126}.

A musical track can be represented as a directed network, whose nodes are the notes played by an instrument. 
When, in a melody, a note $x$ is followed by a subsequent note $y$, a directed link $(x, y)$ from $x$ to $y$ is added to the network. 
This simple mathematical tool can be used to model melodies and musical solos.
Complex networks are an appropriate means to model music as a complex system and provide powerful quantitative measures for capturing the essence of its complexity. 

In a previous work, the approach of modeling melodies has been exploited over a dataset of guitar solos performed by different musicians \cite{Ferretti2017271}.
Using the network representation of these solos, different metrics have been calculated, typical of complex network theory, such as their length of solos, the size and diameter of the networks, the degree distribution, distance metrics, clustering coefficient, centrality betweenness and so on. 
The study revealed that there are statistically significant differences among different artists. This demonstrates the ability of the approach to characterize the main features of an artist or a music genre.
Moreover, the networks analysis showed that most of the considered solos are small worlds. 

The result that different musicians do have different characteristics, in terms of the networks ``they build'' while playing a solo, leads to another question: can these networks/solos be categorized and grouped, so as to understand whose solos are more similar? In this work, we try to deal with this issue.
A clustering technique is employed to group solos into multiple groups, so that solos within a cluster have high similarity, 
while solos in different clusters are dissimilar. 
In particular, the $k$-means clustering technique has been utilized. In order to measure the similarity between two networks, we employed the values of the main metrics that can be derived from the analysis of the networks. These values compose a tuple (characterizing the network) and a classic Euclidean distance is employed.

First of all, using the Hopkins statistic we observe that a clustering tendency exists.
Then, we varied the value of $k$ in the $k$-means clustering algorithm. This value represents the amount of clusters to be used to partitionate the dataset. In such context, the number of clusters is unknown. Through the elbow method we find that, in the employed dataset, a reasonable amount of clusters is in the range of $15-20$ clusters. 

An analysis has been performed on those solos that appear to be in the same cluster, also when varying the number of $k$.
We find that many solos of the same musicians are in the same clusters. This was expected, as each musician has its own style, that would influence the shape and structure of the related networks. Another result is that similar artists (in terms of musicological aspects) do generate similar solos. This is another expected outcome.
Nevertheless, there are also solos, appearing to be similar based on this methodology, which relate to artists that are usually considered playing different genres.
In fact, a clustering based on the main metrics of complex networks
allows grouping melodies based on their general structure, regardless of other musical aspects, such as the instrument that has been utilized to produce them and the specific sound of the instrument. 


\section{Musical Tracks as Networks}\label{sec:musicnet}

Starting from a music sheet, a corresponding network can be built as follows. Nodes of the network correspond to specific notes. The note can be a single one, a rest or a chord, i.e., a group of notes played simultaneously. Each node has a label associated to it.
Labels vary depending on the type of the note.
In case of a single note, the related node has a label composed of the note pitch, octave and duration. A ``rest node'' is labeled with the duration of the rest. Finally, nodes corresponding to chords are labeled with the pitch, octave and duration of each note composing the chord.
Links are associated to nodes that correspond to notes played in sequence in the sheet.

\begin{figure}
\centering
  \includegraphics[width=.6\linewidth]{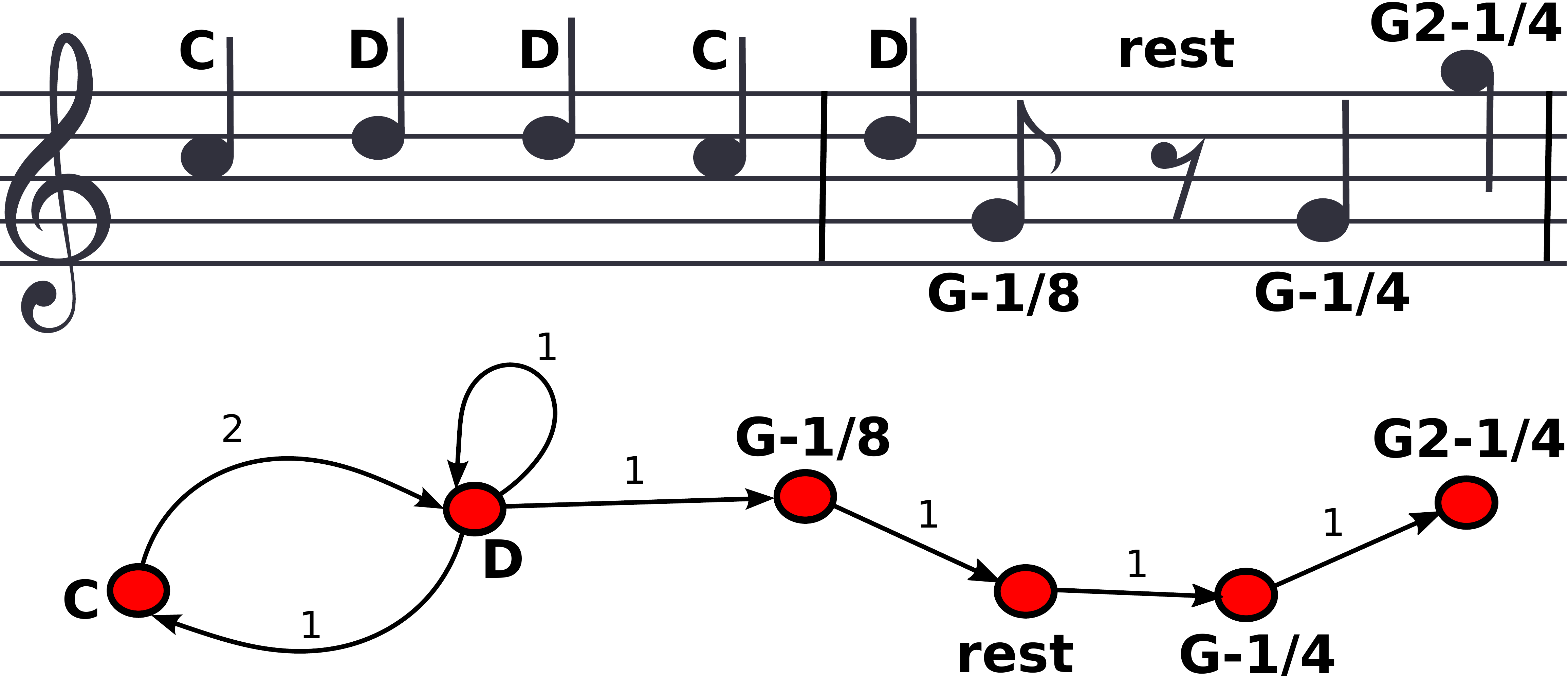}
\caption{Example of melodic line mapped to a network}
\label{fig:netExample}
\end{figure}

Let consider the example shown in Figure \ref{fig:netExample}, where a simple music score sheet of a melodic line is depicted together with the correspondent network. The text label reported over each note is the name of the note. Different notes are mapped into different nodes, that are labeled with the note name. Weights are assigned to links, counting the amount of occurrences of the corresponding notes pair in the score sheet. Links are directed.
Thus, according to the score sheet, a link $(C, D)$ is created from the node $C$ to $D$, since the first note on the sheet is a $C$, followed by a $D$. Then, a self loop $(D, D)$ is added to the network, since the third note on the sheet is a $D$, again. The fourth note is a $C$, that corresponds to the link $(D, C)$. 
A second occurrence of the $(C, D)$ pair increases the weight associated to that link.
Then, there is a sequence of links $(D, G-1/8), (G-1/8, \text{rest}), (\text{rest}, G-1/4), (G-1/4, G2-1/4)$. Note that there are three different nodes for the $G$ notes, since $G-1/8, G-1/4$ have the same pitch (i.e.~$G$) but different duration (the first $G$ is a eighth note, while the second one is a quarter); moreover, $G2-1/4$ is an octave higher than other two $G$ notes.

As a final remark, we notice that nodes and links might be enriched with further information related to specific musical aspects, e.g., a ``legato'' sequence or better, the percentage of links that derive from legato notes. However, in this study we do not consider these additional features.

\begin{figure*}
\centering
\begin{subfigure}{0.55\textwidth}
   \centering
   \includegraphics[width=\linewidth]{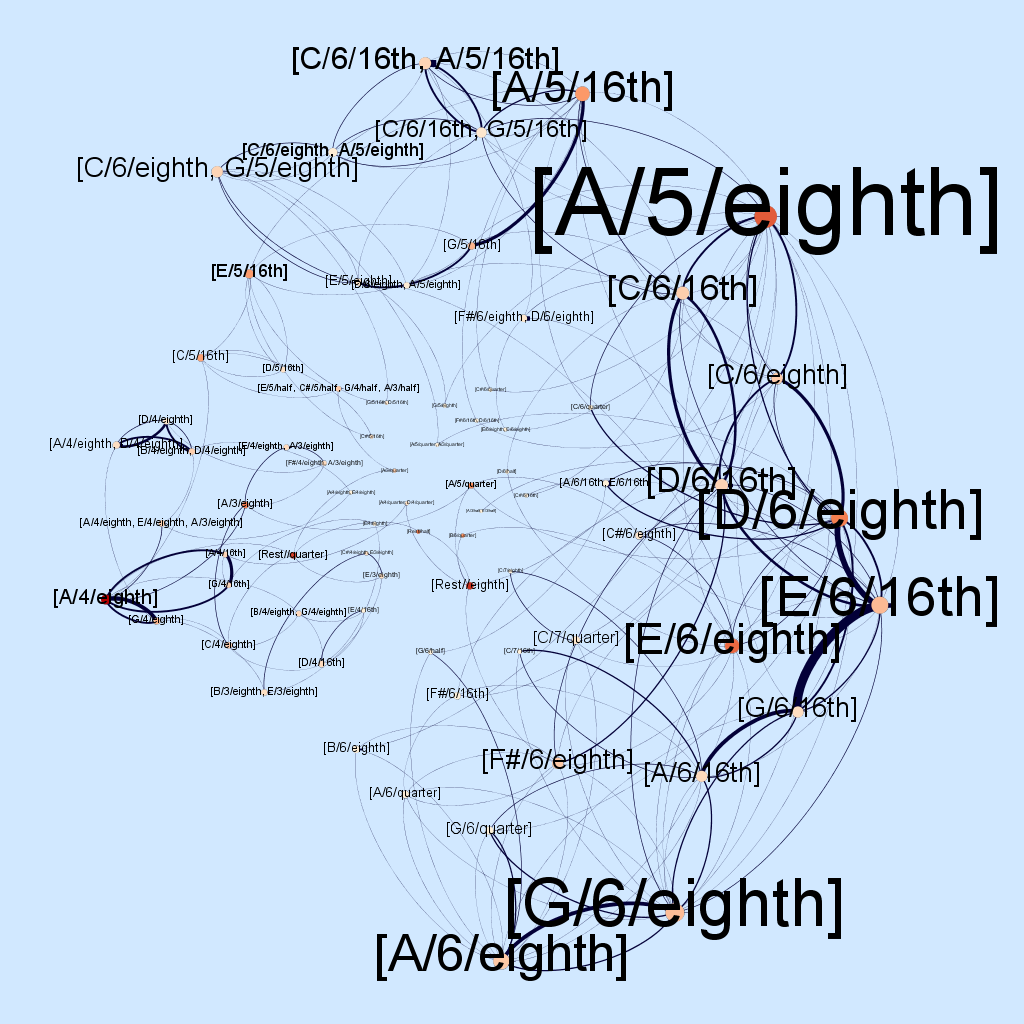}
   \caption{network}
\end{subfigure}%
\begin{subfigure}{0.4\textwidth}
   \centering
  \includegraphics[width=\linewidth]{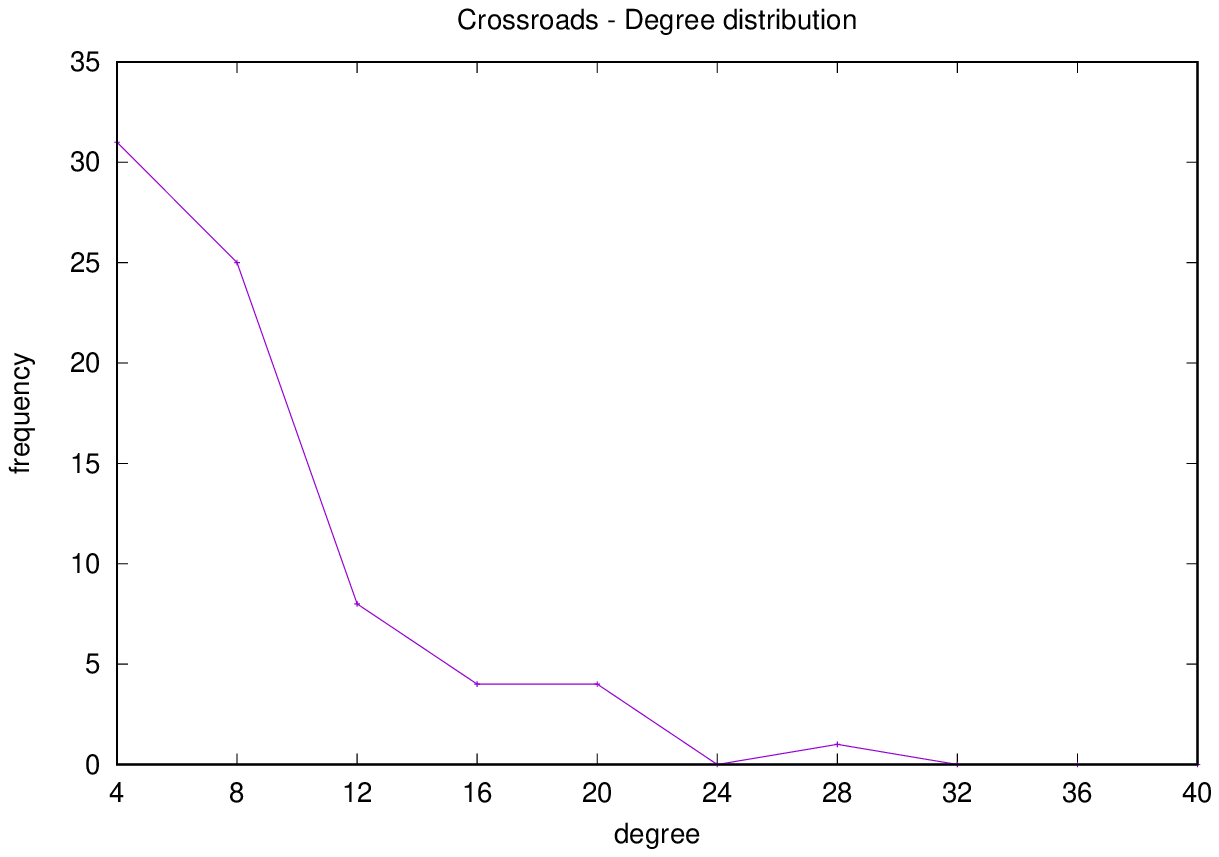}
   \caption{degree distribution}
\end{subfigure}%
\caption{Cream (Eric Clapton) -- Crossroads (2nd solo)}
\label{fig:hendrix}
\end{figure*}

Figure \ref{fig:hendrix} shows the network of the solo played by Eric Clapton in his famous version of the song titled ``Crossroads'', played with the Cream group in the late of the 60s. The left part of the figure is the network, while the chart on the right shows the degree distribution\footnote{The ``degree'' of a node is the number of (incoming or outgoing) links. When weights are associated to links, then the sum of these weights is referred as ``weighted degree''.}.
In the network, nodes have labels associated of the form ``[pitch/octave/du\-ra\-tion]'' where, as the names say, ``pitch'' is the pitch of the sound, ``octave'' is a number counting the octave of the note, and ``duration'' specifies the relative duration of the note, with respect to a given bar.
Nodes have a size (and label font) that is proportional to the degree of the node, i.e., the higher the amount of connections the larger the node (and label) size. 

The color of nodes is proportional to the measure of the betweenness centrality measure; the more the color goes to red the higher the betweenness of that node \cite{Ferretti2017271}. In short the betweenness value of a node $x$ states how much that node is central in the network, i.e.,~how frequent is that, going from a note to another, the player passes through that note $x$.
In summary, nodes with a red color and a larger size reflect some important, common notes in the melodic line. In this case, for example, it is clear that in this solo the main note is [A/5/eighth]. Indeed, the song is a blues played in the keynote of A.
Links are depicted based on their weight, i.e., the larger the line the higher the weight of that link in the network, meaning that the notes' pair connected through that link has been played multiple times. In this representation, to simplify the graphic appearance of the network, links are considered as undirected.
Thus, by looking at the size of the links, it appears evident that in this solo the player quite frequently played the note sequences ([E/6/16th], [G/6/16th]) (or vice versa) and ([E/6/16th], [D/6/eighth]) (or vice versa).

The degree distribution shows that while the majority of nodes has a low degree, there are some main nodes that act as hubs for the network, since their degree is higher than others. 

From this network representation, it is possible to measure all the typical metrics associated to a network. For instance, the average degree of the network is $3.2$, the diameter is $12$, the average path length is $5.17$, its clustering coefficient is $0.25$ (please note that this coefficient is a network specific parameter, and it has nothing to do with the clustering of different networks performed in the following of this paper).

\section{Background: Clustering of Musical Networks}\label{sec:clustering}

Clustering is the process of grouping a set of data objects into multiple groups, so that objects within a cluster have high similarity, 
while objects in other clusters are dissimilar. The similarity is assessed based on some parameters to be chosen. 
Clustering musical tracks would allow identifying their similar aspects. In particular, when applied to melodies, it is possible to group them based on their general structure, regardless of other musical aspects, such as the instrument that has been utilized to produce them and the specific sound of the instrument. 

An important first question is whether there is a real tendency of these musical tracks to cluster together. Then, once a clustering algorithm is utilized, it is important to understand if the resulting clusters are good ones. In this section we discuss these issues, while next section will show results of a clustering algorithm applied to a database of musical melodies.

In order to measure the similarity between two networks, we employed the values of the main metrics that can be derived from the analysis of the networks. These values compose a tuple characterizing the network, and a classic Euclidean distance is employed.

\subsection{Assessing clustering tendency}

Assessing clustering tendency is an important task, since it allows understanding whether a
non-random structure exists in the considered dataset. In fact, clustering analysis on a data set is meaningful only when there is a non-random structure in
the data \cite{Han:2005}.
To this aim, a common procedure is related to the calculation of the Hopkins statistic. It examines whether objects in a data set differ significantly from the assumption that they are uniformly distributed in a multidimensional space. 

To this aim, a set of artificial random points is uniformly generated in the data space where the real data lie.
Then, for each real data object $i$ in the dataset, we measure the distance $r_i$ between $i$ and its nearest neighbor. The obtained value is compared with a similar measure made on the set of generated artificial objects. Thus, for each artificial object $j$, we measure the distance $a_j$ between $j$ and its nearest real neighbor (in this case, the label $r$ identifies for the distances of ``real'' objects, while $a$ identifies for the distances of ``artificial'' ones). 
The process is repeated for a fraction (or the whole set) of the real objects. The Hopkins statistic is measured as
$$H = \frac{\sum_i a_i}{\sum_i r_i + \sum_i a_i}.$$

If real data objects are uniformly distributed, the two summations of $r_i$ and $a_i$ values will be similar; thus $H$ will be close to 0.5. Conversely, when a
clustering is present, the distances for artificial objects will be larger than those for the real ones, because these artificial objects are homogeneously 
distributed whereas the real ones are grouped together, and the value of $H$ increases. Thus, if $H > 0.5$ we have a clustering tendency.

\subsection{The clustering approach}

The clustering approach that has been utilized is the well-known partitioning algorithm named $k$-means \cite{Han:2005}. The algorithm partitionates the dataset into $k$ clusters. During the tests, we varied the value of $k$.

Outcomes of the clustering algorithm execution were evaluated by measuring the Sum of Squared Error (SSE) and the silhouette value.
The SSE is the (squared) sum of the distance of each considered object to the centroid of its assigned cluster. It measures the within-cluster variation.

The silhouette value evaluates a clustering by examining how well the clusters are separated and how compact the clusters are. 
It measures the average distance between each object $x$ and all other objects in the cluster to which $x$ belongs. This value is compared to the minimum average distance from the object $x$ to all clusters to which $x$ does not belong. Values are then normalized so that the coefficient value is between $-1$ and $1$. A good clustering is obtained when the silhouette value is near $1$, and higher than $0$ in general.

In order to assess which is the best number of clusters $k$, we can exploit the elbow method applied to the SSE \cite{Han:2005}.
The elbow method is based on the observation that increasing the number of clusters
can help to reduce the sum of within-cluster variance of each cluster. This is because
having more clusters decreases the distance of the objects from their centroids. However, 
above a certain $k$ value, such reduction becomes marginal. 
The elbow method suggests selecting the right $k$ values as the turning point in the curve of the SSE with respect to the number of clusters.

Then, we adopted the following approach to determine which musical tracks can be considered as similar.
Given each track and its corresponding network, its network-related metrics were collected.
This set of parameter values were used as the tuple objects provided as inputs to the $k$-means clustering approach. 
We repeated the execution of this algorithm several times, for each $k$ value. 
We then created a network composed of all the considered solos. Every time two solos appear to be in the same cluster, we added a link between these two nodes (and then, the link weight was increased).
The use of weights allows filtering and removing those links whose weights are below a certain threshold. 
The result was a highly clustered network. In this network, maximal cliques were identified using the Bron-Kerbosch algorithm.


\section{Clustering of Guitar Solos}\label{sec:perf}

The presented approach can be applied to all musical tracks in general. In this work, we focus on a database of $\sim150$ guitar solos. 
Such a database was collected from the amount of music score sheets that is available on the Web, typically encoded according to some guitar-oriented notation systems (e.g.,~guitar tablature). Those who are interested on the details about the database retrieval and data manipulation can refer to \cite{Ferretti2017271}. 
The idea of employing solos is due to the fact that, during a solo, a player creates a melody usually ``in real time''; during this process, it is reasonable to assert that he employs typical patterns (licks) he is used to utilize \cite{barrett,Boothroyd}.

\subsection{Software for the analysis}

The software to perform the analysis was built using Java programming language, mainly. 
An in-house software was developed, that resorts to the open-source software JUNG (Java Universal Network/Graph) Framework to manipulate networks and extract the metrics of interest \cite{jung}.
Moreover, the Apache Commons Mathematics Library was exploited to perform the mathematics and clustering analysis \cite{apache}.
The source code of the software is freely available, upon request.

\subsection{Selection of artists}
The selection of artists was based on three criteria, i.e.,~the ``importance'' of the performer, the amount of songs available in the Web for that performer, the diversity (from a musical point of view) from other performers. 
The idea was to consider a wide range of different musical styles, to see if there are any differences in the corresponding obtained networks. 

We thus chose those musicians that, according to the general music criticism, have a unique playing style. 
Briefly, the artists are: 
\begin{itemize}
 \item Eric Clapton (EC), a rock-blues guitar legend (the slogan ``Clapton is God'' testifies this); 
 \item David Gilmour (DG), singer and guitarist of the famous Pink Floyd rock group, known for his melodic and intense guitar solos; 
 \item Jimi Hendrix (JH), which is considered the most important electric guitar player of all times; 
 \item Allan Holdsworth (AH), a fusion artist noted for his advanced style and his intricate solos;
 \item B.B.~King (BBK), a blues master quite often referred as ``The King of the Blues'', that inspired several generations of musicians; 
 \item Pat Metheny (PM), probably the most famous contemporary jazz guitar player;
 \item Steve Vai (SV), a well known guitar virtuoso, famous for his ``unique'' approach to the instrument;
 \item Eddie Van Halen (EVH), which is considered one of the most influential hard-rock guitarists of the 20th century.
\end{itemize}

\subsection{Parameters exploited during the clustering}

The parameters exploited to perform the clustering were selected among the following ones:
the number of nodes of the network, number of edges, length of the solo, diameter of the network, number of notes per bar, clustering coefficient of the network, degrees statistics (average, median, minimum, maximum of the degrees, in-degrees, weighted degrees), statistics on the betweenness centrality value (average, median, minimum, maximum).

Due to the number of considered parameters, the database results as a high-dimensional data. This may represent a problem, since it is well known that when the dimensionality is high, conventional distance measures can be dominated by noise. In this study we varied the set of considered parameters, thus reducing the dimensionality of the dataset.

\section{Results}\label{sec:results}

\subsection{Clustering tendency of guitar solos}

We already mentioned that, in general, the first thing to do is to assess if a database of objects (in this case, musical tracks containing guitar solos) has a clustering tendency. To this aim, we exploited the already mentioned Hopkins statistic.
We made $20$ tests over the considered database, obtaining an average value of $H = 0.83$. This confirms a clustering tendency on the set of considered solos.

\subsection{Clustering outcomes}

The adopted approach is quite resilient to the variability of the parameter $k$. In fact, some sub-clusters are present in every different instance of the execution of the clustering algorithm, regardless of the $k$ parameter. This suggests that these solos do have similar traits. 

\begin{figure*}
\centering
  \includegraphics[width=.9\linewidth]{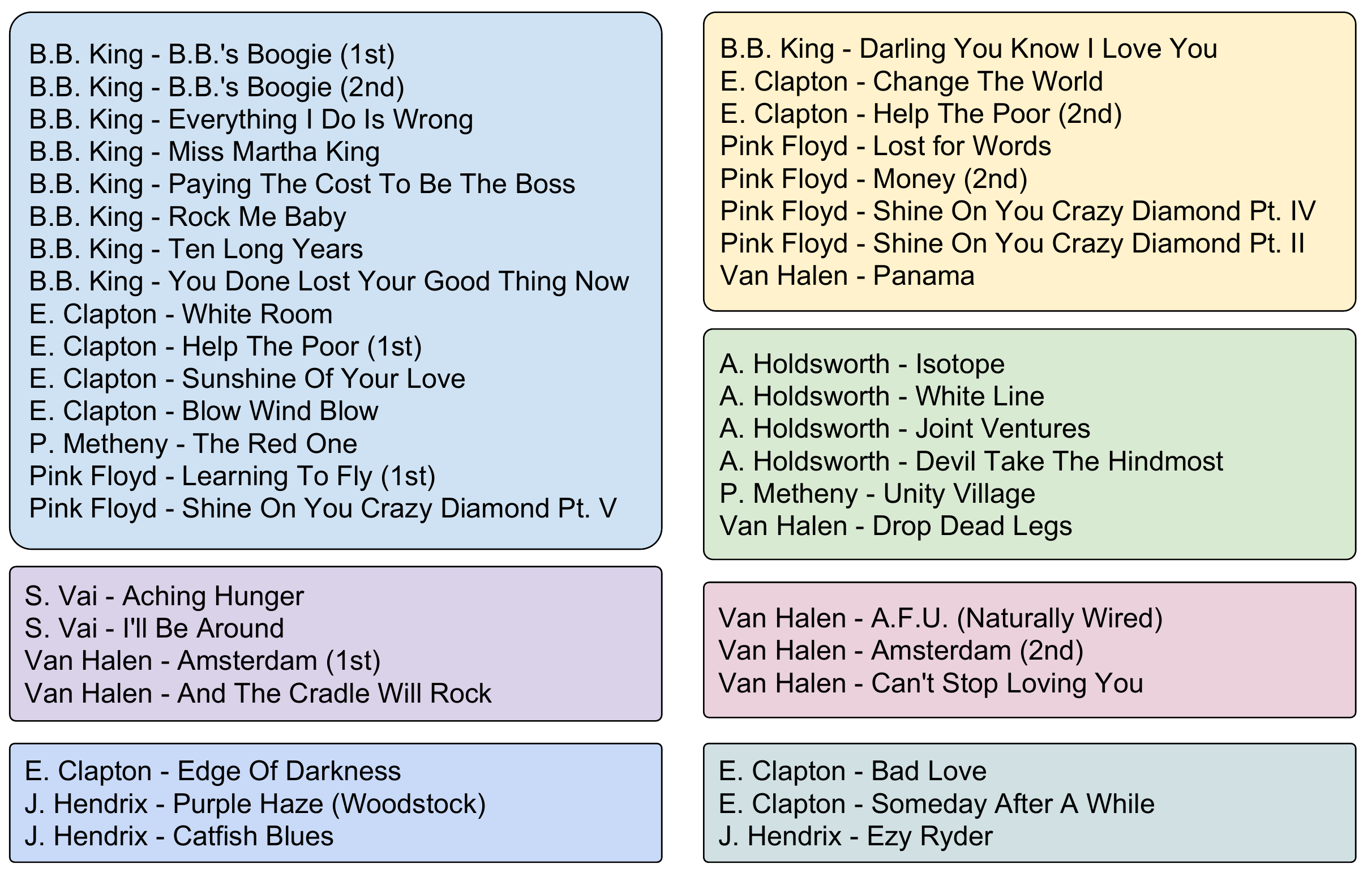}
\caption{Some clusters of solos; clusters are groups of solos inside boxes.}
\label{fig:clusters}
\end{figure*}

Figure \ref{fig:clusters} shows a set of selected clusters of solos (each box is a cluster).
It is interesting to notice that there are several solos of same musicians clustered together. This confirms the ability of the approach to group solos which are correlated, since each artist has his own style to create a melody.
Moreover, the figure reports examples of solos from artists which are considered similar from a musical point of view, since they play similar musical genres. Examples are B.B.~King and E.~Clapton, E.~Clapton and D.~Gilmour (Pink Floyd), S.~Vai and E.~Van Halen.

However, an interesting outcome is that, actually, we also obtain some similarity among solos from artists that are usually considered very different, from a musical point of view. In other words, there are artists whose music can be categorized in different music genres, whose solos have something in common, nevertheless.
To explain this results, it is worth noticing that the approach compares scores, which are composed of information that does not take into consideration aspects such as the underlying backing track (that can be very diverse in different songs) and the ``sound'' of the guitars. For example, a player might use a cleaner sound while another can exploit a distorted one with other sound effects; these two settings would make the same note sound very different. Conversely, in this model the note is treated as a note, and the combination of these notes is not altered by the sound used to play these notes.

This last result, that can be considered as counter-intuitive at first glance, is in line with the assertion that most musicians have been influenced that other ones, quite often coming from different music genres. For instance, E.~Van Halen confirmed that E.~Clapton and J. Hendrix had a strong influence in his playing, even if their musical genres (and their guitar sounds) are different.

\begin{figure}
\centering
  \includegraphics[width=.7\linewidth]{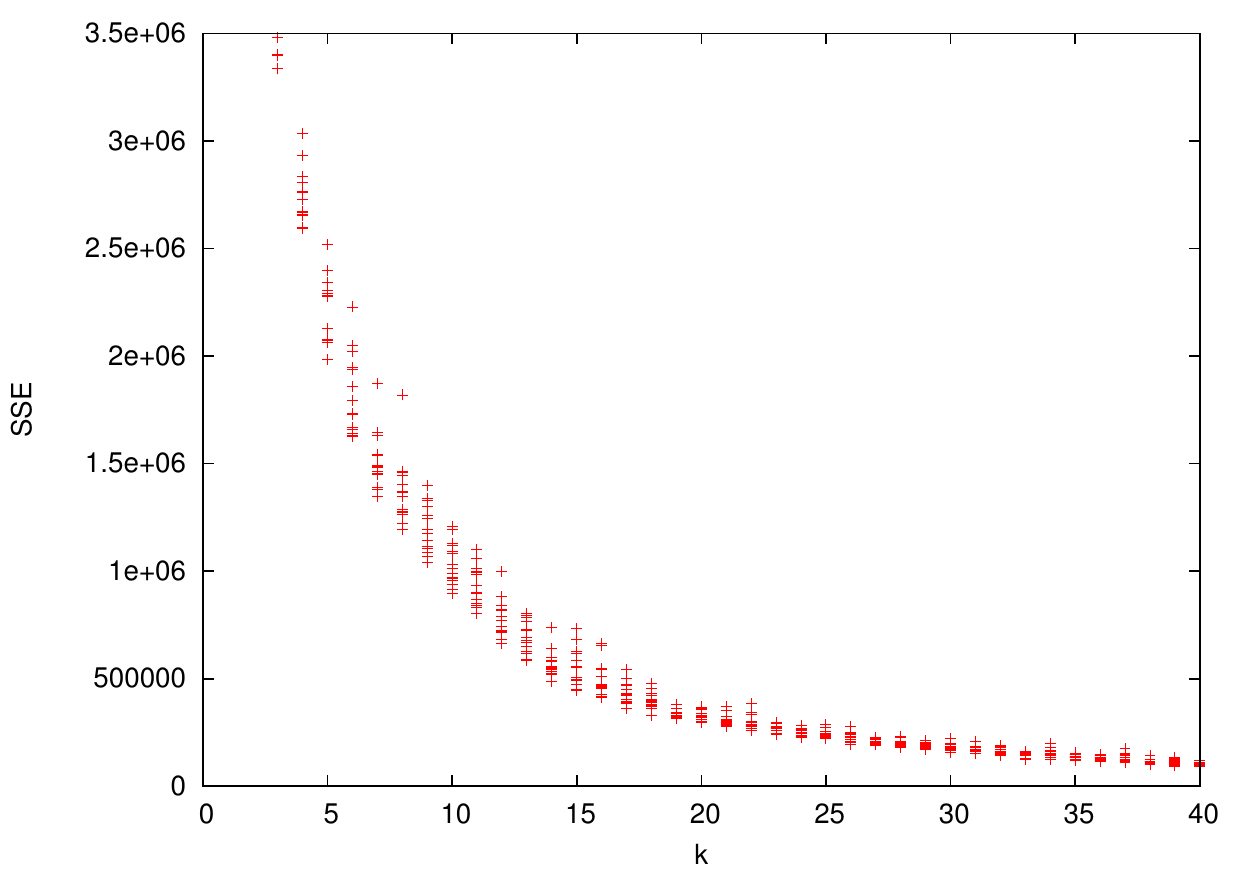}
\caption{SSE obtained for different values of the K-means algorithm}
\label{fig:sse}
\end{figure}

As concerns the choice of the number of clusters $k$, we exploited the elbow method by plotting the Sum of Squared Error (SSE), depending on the number of clusters $k$.
Figure \ref{fig:sse} shows such variation of the SSE for this considered dataset. Following this approach, 
the turning point in the curve of the SSE, with respect to the number of clusters $k$, lies around $15-20$ clusters. Thus, the number of clusters $k$ should be selected in this range.

Depending on the selection of the parameters and the value of $k$, the silhouette value can vary from $\sim 0.4$ up to $\sim 0.6$.
This demonstrates that the approach is able to properly identify good clusters.

%
%
%
%
%
%
%
%
%
%
%
%

\subsection{Similarity among authors}

\begin{table*}[th]
\caption{Percentage of tracks of different artists appearing in the same clusters.} 
\begin{center}
\begin{tabular}{ | c || c| c | c | c | c | c | c | c |  }
\hline
	& JH   & EC   & DG   & EVH   & AH   & SV   & BBK   & PM	\\ \hline
\hline
JH      & 0.28 &      &      &       &      &      &	   &		\\ \hline
EC	& 0.32 & 0.35 &      &       &      &      &	   &		\\ \hline
DG	& 0.27 & 0.35 & 0.3  &       &      &      &	   &		\\ \hline
EVH	& 0.31 & 0.28 & 0.29 & 0.37  &      &      &	   &		\\ \hline
AH	& 0.13 & 0.19 & 0.14 & 0.13  & 0.23 &      &	   &		\\ \hline
SV	& 0.2  & 0.13 & 0.11 & 0.2   & 0.12 & 0.16 &	   &		\\ \hline
BBK	& 0.21 & 0.37 & 0.39 & 0.12  & 0.08 & 0.03 & 0.72  &		\\ \hline
PM	& 0.12 & 0.24 & 0.19 & 0.1   & 0.31 & 0.05 & 0.21  & 0.39	\\ \hline
\end{tabular}
\end{center}
\label{table:authors}
\end{table*}

Table \ref{table:authors} shows the percentage of tracks of a given artist that are clustered to some other track of another artist. The value is normalized with respect to the amount of possibilities. Thus, if we consider two different artists $x$ and $y$, the reported value is the amount of tracks of $x$ and $y$ appearing in the same clusters (or put in a network perspective, the amount of links between tracks of $x$ and those of $y$) divided by the number of possible combinations of tracks of $x$ and $y$, i.e.,~$(|\textnormal{tracks}_x| \cdot |\textnormal{tracks}_y|)$, where $|\textnormal{tracks}_x|$ is the number of tracks of artist $x$ (and the same holds for $y$).
Conversely, if we consider tracks of the same artist $x$, the value shown in the table is the amount of tracks of $x$ appearing in the same clusters, divided by all the possible combinations among these tracks (i.e.,~$|\textnormal{tracks}_x| \cdot (|\textnormal{tracks}_x|-1)/2$).
We report only the lower triangular part of the matrix, since the matrix is symmetric.
Based on the previous discussion, concerned with the amount of $k$ clusters to employ in the clustering algorithm (Figure \ref{fig:sse}),
results relate to the $k$-means algorithm with $k=17$.

The idea behind this study is that, since the dataset is composed of an amount of tracks coming from a limited set of artists, and since each artist has his own characteristics, from a musicological point of view assessing how many tracks of the same artist (or similar artists) are clustered together can be an interesting outcome to evaluate the clustering technique.

Indeed, the table provides interesting results. For instance, it can be seen that a high portion of songs by B.B.~King ($0.72$ in Table \ref{table:authors}) are in the same clusters. This confirms the general understanding that this musician was used to play similar licks (i.e.,~patterns and motifs) in his solos (even if, from a musical point of view, these licks are considered as very influential in the blues domain). It is also possible to observe a high similarity between B.B.~King and Eric Clapton, as well as the guitar player leading the Pink Floyd group, David Gilmour. Indeed, Clapton is another prominent blues player, while it is recognized that Gilmour has a melodic playing-style, which can be somehow compared to that of B.B.~King. 
Conversely, B.B.~King has a low intersection with Steve Vai and Allan Holdsworth, which have playing styles that are very different from him.

Other significant percentages of clustered tracks are those coming from 
Pat Metheny and himself, 
Eddie Van Halen and himself, 
Eric Clapton and himself, 
Jimi Hendrix and Eric Clapton,
Pat Metheny and Allan Holdsworth.

These results, coming from the clustering algorithm, are another confirmation that the employed approach is a good methodology to model music tracks and melodies. Actually, it is not possible to compare such outcome with other quantitative measures. Thus, these numerical results are to be combined with the musicological studies that assess aspects related to improvisation and musical styles \cite{covach1997understanding,Pressing,smith,zenni}.

\section{Conclusions}
\label{sec:conc}

In this work, we performed a clustering analysis on a dataset of guitar solos. Such musical melodies have been modeled as networks. Main complex networks metrics have been measured and utilized as parameters for the clustering.

As a first result, we noticed that a clustering tendency exists, through the measurement of the Hopkins statistic. 
Then, we applied the clustering technique and identified a viable range of number of clusters to be employed. 
The resulting groups contain tracks coming from the same artists, or similar ones. 
Conversely, we noticed a limited amount of ``inter-linked'' tracks of artists which are considered very dissimilar from a musicological point of view.
This is a confirmation of the viability of the proposal. 

There is a number of possible improvements and future works. 
The employed clustering scheme was a $k$-means scheme with an Euclidean distance exploited to measure the distance among data objects. An approach might be that of employing a different clustering scheme and some more sophisticated distance. 

Moreover, the amount of parameters employed to characterize each track was high. 
In this sense, it has been recognized that the clustering of high-dimensional data might result as a complicate task.
During the study, we varied and limited the number of these parameters. However, the use of some specific approach, thought to be used over high-dimensional data, such as subspace clustering approaches or dimensionality reduction approaches, might provide better results.


\bibliographystyle{elsart-num-sort}
\bibliography{biblio}  


\end{document}